\def\lsim{\mathrel{\rlap{\lower4pt\hbox{\hskip1pt$\sim$}}
    \raise1pt\hbox{$<$}}}         
\def\gsim{\mathrel{\rlap{\lower4pt\hbox{\hskip1pt$\sim$}}
    \raise1pt\hbox{$>$}}}         
\def\Imag#1{\Im{\rm m}#1}
\def\Real#1{\Re{\rm e}#1}

\def\ie{\hbox{\it i.e. }}

\def\eg{\hbox{\it e.g. }}

\def\etal{\hbox{\it et al. }}

\def\g2{ GeV$^2$}
\def\xi2{$\chi^2_{d.o.f}$}
\def\underarrow#1{\mathrel{\mathop{\longrightarrow}\limits_{#1}}}

\documentclass[12pt]{article}
\usepackage{graphicx}
\begin{document}

\rightline{LYCEN 9906}
\rightline{february 10}

\bigskip

\begin{center}

{\bf {\Large {UNITARITY, (ANTI)SHADOWING 
AND THE BLACK DISC LIMIT}}}

\bigskip

\large{
P. Desgrolard (\footnote{E-mail: desgrolard@ipnl.in2p3.fr}),
L. Jenkovszky (\footnote{E-mail: jenk@bitp.kiev.ua}).
B. Struminsky (\footnote{E-mail: eppaitp@bitp.kiev.ua})}

\end{center}

\bigskip

\noindent
($^1$) {\it Institut de Physique Nucl\'eaire de Lyon, IN2P3-CNRS et
Universit\'e Claude Bernard,
43 boulevard du 11 novembre 1918, F-69622 Villeurbanne Cedex, France}

\noindent
($^{2, 3}$) {\it Bogolyubov Institute for Theoretical Physics,
National Academy of Sciences of Ukraine, 252143 Kiev, Ukraine.}

\bigskip
\noindent
{\bf \large Abstract}
By using realistic models for elastic hadron scattering we
demonstrate that at present accelerator energies the $s$-channel
unitarity bound is safe, not to be reached
 until $10^5$ GeV, while
the black disc limit is saturated around 1 TeV.  It will be followed
by a larger transparency of the scattered particles near the center.

\bigskip

\section{Introduction}

Our decision to write this paper was motivated partly by recent claims
that in high energy hadron scattering the black disc limit has been reached
and the violation of the $s$-channel unitarity in some models is
just around the corner. While the first statement is true and has
interesting physical consequences, the second one is wrong for any
realistic model fitting the existing data on proton and antiproton
scattering up to highest accelerator energies.

To start with, let us remind the general definitions and notations.

\noindent
Unitarity in the impact parameter ($b$) representation reads
\begin{equation}
\Imag {h(s,b)}=\big|h(s,b)\big|^2+G_{in}(s,b),
\end{equation}
where $h(s,b)$ is the elastic scattering amplitude
at $\sqrt{s}$ center of mass energy
(with $\Imag h(s,b)$ usually called the profile function,
representing the hadron opacity) and
$G_{in}(s,b)$, called the inelastic overlap function, is the sum
over all inelastic channel contributions.
Integrated over ${b}$,
(1) reduces to a simple relation between the total, elastic and
inelastic cross sections
$\sigma_{tot}(s)=\sigma_{el}(s)+\sigma_{in}(s).$

Equation (1) imposes an absolute limit
\begin{equation}
0 \leq \big|h(s,b)\big|^2\leq \Imag h(s,b) \leq 1,
\end{equation}
while the so-called "black disc" limit
$\sigma_{el}(s)=\sigma_{in}(s)={1\over 2}\sigma_{tot}(s)$
or
\begin{equation}
\Imag h(s,b) = 1/2
\end{equation}
is a particular realization of the optical model, namely it
corresponds to the maximal absorption within the eikonal
unitarization, when the scattering amplitude is approximated as
\begin{equation}
h(s,b)={i\over 2}\big(1-\rm{exp}\left[i\omega(s,b)\right]\big),
\end{equation}
with a purely imaginary eikonal $\omega(s,b)=i\Omega(s,b)$.

Eikonal unitarization corresponds to a particular solution of the
unitarity equation
\begin{equation}
h(s,b)={1\over 2}\left[1\pm\sqrt{1-4G_{in}(s,b)}\right],
\end{equation}
the one with minus sign.

The alternative solution, that with plus sign is known
\cite{TT,BP} and realized within the so-called $U$-matrix
\footnote{We follow traditional terminology, although the word
"matrix" in this context is misleading, since $U$, similar to the
eikonal, is a single function rather than a matrix.} approach
\cite{umatrix,VJS} where the unitarized amplitude is
\begin{equation}
h(s,b)={\Imag{U(s,b)}\over{1-i\ U(s,b)}},
\end{equation}
where now $U$ is the input "Born term", the analogue of the eikonal
$\omega$ in (4).

In the $U$-matrix approach, the scattering amplitude $h(s,b)$ may
exceed the black disc limit as the energy increases. The
transition from a (central) "black disc" to a (peripheral) "black ring",
surrounding a gray disc, for the inelastic overlap
function in the impact parameter space corresponds to the
transition from shadowing to antishadowing \cite{TT}. We shall
present a particular realization of this regime.

The impact parameter amplitude may be calculated either directly from
the data, as it was done \eg in \cite{ASch,CS} (where,
however, the real part of the amplitude was neglected) or by using a
particular model that fits the data sufficiently well. There are
several models appropriate for this purpose. In a classical paper
\cite{BEL} on this subject, from the behaviour of $G_{in}(s,b)$,
the proton is characterized as getting "BEL" (Blacker, Edgier and
Larger). As anticipated in the title of our paper, the proton,
after having reached its maximal darkness around the Tevatron
energy region, may get less opaque beyond.

Actually, the construction of any scattering amplitude rests on
two premises~: the choice of the input, or "Born term" and the
relevant unitarization procedure (eikonal or $U$-matrix in our case).
Within the present accelerator energy region there are several
models that fit the data reasonably well. Compatible within the region of
the present experiments, they differ significantly when extrapolated
to higher energies.
We shall
consider two representative examples, namely the Donnachie-Landshoff (D-L)
model \cite{DL,DL1} and the dipole Pomeron (DP) model \cite{VJS,DGJ})

In Sec.~2 we present the necessary details about the two realistic 
models (D-L and DP), then, focusing on the DP model, we investigate in 
Sec.~3 the unitarity properties at the "Born level" and in Sec.~4 we 
study the optical properties (transparency) after unitarization; a 
comparaison with the D-L model is given in the Appendix.

\bigskip

\section{The "Born term"}

The Donnachie-Landshoff (D-L) model \cite{DL} is popular for its
simplicity.  Essentially, it means the following four-parametric
empirical fit to all total hadronic cross sections
\begin{equation}
\sigma_{tot}=X\ s^{\delta}+Y\ s^{\delta_r},
\end{equation}
where two of the parameters, namely  $\delta=\alpha_P(0)-1\approx 
0.08$, and $\delta_r (<0)$ are universal. While the violation of the 
Froissart-Martin (F-M) bound, \begin{equation} \sigma_{tot}(s) < C\ 
(\ell n {s})^2\ \quad C=60\ {\rm mb}\ , \end{equation} inherent in 
that model, is rather an aesthetic than a practical defect (because 
of the remoteness of the energy where eventually it will overshoot 
the F-M limit), other deficiencies of the D-L model (or any other 
model based on a supercritical Pomeron) are sometimes criticized in 
the literature, but so far nobody was able to suggest anything 
significantly better instead. A particular attractive feature of the 
D-L Pomeron, made of a single term, is its factorizability, although 
this may be too crude an approximation to reality.

The $t$ dependence in the Donnachie-Landshoff model is usually chosen
\cite{DL1} in the form close to the dipole formfactor. For the
present purposes a simple exponential residue in the Pomeron amplitude will do
as well, with the signature included
\begin{equation}
A(s,t)= -\ N \left(-i{s\over s_{dl}}\right)^{\alpha(t)} e^{Bt}\ ,
\end{equation}
where $\alpha(t)=\alpha(0)+\alpha'\ t $ is the Pomeron trajectory and
$N$ is a dimensionless normalization
factor related to the total cross section at $s=s_{dl}$ by the optical theorem
\begin{equation}
N = {s_{dl}\over 4\pi\sin{{\pi\over 2}\alpha(0)}}\sigma_{tot}(s= s_{dl})\ .
\end{equation}

\noindent
According to the original fits \cite{DL,DL1}: $s_{dl}=1$ GeV$^2$,
$\alpha(0)=1.08$, $\alpha'=0.25$ GeV$^{-2}$, and $X=21.70$ mb (see 
(7)) resulting in $N={X\over 4\pi\sin{\pi\alpha(0)/2}}=4.44$. By 
identifying \begin{equation} {d\sigma(s,t)\over dt}\ 
={d\sigma(s,t=0)\over dt}\ e^{B_{exp}(s)\ t} \end{equation} and 
choosing the CDF or E410 result for the slope $B_{exp}$ at the 
Tevatron energy, we obtain $B= {1\over 2}B_{exp}(s) -
\alpha'\ell n{{s\over s_{dl}}}=4.75$ GeV$^{-2}.$
\medskip

In the dipole Pomeron (DP) model \cite{VJS},
factorizable  at asymptotically high energies,
logarithmically rising cross sections are produced at a unit
Pomeron intercept only and thus DP does not conflict with the F-M bound.
While data on total cross section are compatible with a logarithmic
rise (DP with unit intercept) the ratio $\sigma_{el}/\sigma_{tot}$ is found (see
\cite{Yaf} for details) for $\delta=0$ to be a monotonically decreasing function of
the energy for any physical value of the parameters. The
experimentally observed rise of this ratio can be achieved only for
$\delta > 0$ and thus requires the introduction of a "supercritical"
Pomeron, $\alpha(0)>1$. As a result, the rise of the total cross sections is
driven and shared by the dipole and the "supercritical" intercept.
The parameter $\delta=\alpha(0)-1$ in the DP model is nearly half that
of the D-L model, making it safer from the point of view of the
unitarity bounds. Generally speaking, the closer the input to the unitarized
output, the better the convergence of the unitarization procedure.

Let us remind that apart from the "conservative" F-M bound, any model
should satisfy also $s$-channel unitarity. We demonstrate below that both
the D-L and DP model are well below this limit and will remain so within
the forseable future.
(Let us remind that the D-L and the DP model are close numerically, 
although they are different conceptually and consequently they 
extrapolations to superhigh energies will differ as well.)

The elastic scattering amplitude corresponding to the
exchange of a dipole Pomeron reads
\begin{equation}
\begin{array}{rcl}
A(s,t)=&{d\over{d\alpha}} \left[{\rm
e}^{-i\pi\alpha/2}G(\alpha)(s/s_0)^\alpha\right]\\ =& {\rm
e}^{-i\pi\alpha/2}(s/s_0)^\alpha
\big[G'(\alpha)+(L-i\pi/2)G(\alpha)\big]\ ,
\end{array}
\end{equation}
where $L\equiv\displaystyle{\ell n{s\over{s_0}}}$, $\alpha\equiv\alpha(t)$
is the Pomeron trajectory; in this paper, for simplicity we use a linear
trajectory
$\alpha(t) = \alpha(0)+\alpha' t$.

By identifying $G'(\alpha)=-a{\rm e}^{b_p(\alpha-1)},$
(12) can be rewritten in the following "geometrical" form
\begin{equation}
A(s,t)=i{{as}\over{b_p s_0}}\left[r_1^2(s)\ {\rm
e}^{r^2_1(s)[\alpha(t)-1]}- \epsilon\ r_2^2(s)\ {\rm
e}^{r_2^2(s)[\alpha(t)-1]}\right]\ ,
\end{equation}
where
\begin{equation}
r^2_1(s)=b_p+L-i{\pi\over 2}\ , \ r_2^2(s)=L-i{\pi\over 2} \ .
\end{equation}
The model contains the following adjustable parameters:
$a, b_p, \alpha(0), \alpha',\epsilon$ and $s_0$.

In Table 1 we quote the numerical values of the parameters of the
dipole Pomeron
fitted in \cite{DGJ} to the data on proton-proton and
proton-antiproton elastic scattering~:
\begin{equation}
\sigma_{tot}(s)={4\pi\over s} \Imag A(s,0)\ ,
\ \rho (s)={\Real A(s,0)\over{\Imag A(s,0)}}
\quad ; \qquad 4 \leq \sqrt s ({\rm GeV}) \leq 1800
\end{equation}
as well as the differential cross-section
\begin{equation}
{d\sigma(s,t)\over dt}\ =\ {\pi\over s^2}\left|A(s,t)\right|^2 \ ;
\quad   23.5\leq \sqrt s ({\rm GeV}) \leq 630 \ ;
\quad 0 \leq |t|({\rm GeV}^2) \leq 6\ .
\end{equation}
In that fit, apart from the Pomeron, the Odderon and two subleading
trajectories $\omega$ and $f$ were also included. Here, for
simplicity and clarity we consider only the dominant term at high energy
due to the Pomeron
exchange  with the parameters fitted in \cite{DGJ}.
The extent to which this Pomeron is a good approximation in the TeV region is
discussed in details in \cite{Chatr}.

\begin{table}
\caption{Parameters of the Dipole Pomeron found in \cite{DGJ}.}
\medskip
\begin{center}
\begin{tabular}{|c|c|c|c|c|c|}
\hline
$a$ & $b_p$ & $\alpha(0)$ & $\alpha'$(GeV$^{-2}$)  & $\epsilon$ &
$s_0$(GeV$^2$)\\
\hline
355.6 & 10.76 & 1.0356 & 0.377 & 0.0109 & 100.0\\
\hline
\end{tabular}
\end{center}
\end{table}

The quality of this fit is illustrated and discussed in \cite {DGJ}.
With such a simple
model and small number of parameters, better fits are hardly to be
expected.

We use the above set of parameters to calculate the impact parameter
amplitude, and to scrutinize in Sec.~3  the unitarity
properties of this "Born level" amplitude. In Sec.~4 we introduce a
unitarization procedure, necessary at higher energies and discuss
the relevant physical consequences.

To summarize, the DP model with a unit intercept is selfconsistent in
the sense that its functional (logarithmic) form is stable with
respect to unitarization. Moreover, the presence of the second term,
proportional to $\epsilon$ in (13) has the meaning of absorptions
and it is essential for the dip mechanism. It can be viewed also as
one more unitarity feature of the model.

\noindent
In the limit of very high energies, when $L \gg b_p$
the two (squared) radii $R^2_i=\alpha'r^2_i$ become equal and real
and the model obeys exact geometrical scaling as well as factorization (see
next section). Alternatively, it corresponds to the case of no
absorptions ($\epsilon=0$).

\noindent
However attractive, the case of a unit intercept ($\delta=0$)
is only an approximation to the more realistic model, requiring
$\delta>0$ to meet the observed rise of the ratio $\sigma_{el}/\sigma_{tot}$.
For such a "supercritical" Pomeron unitarization becomes inevitable.

\bigskip

\section{Impact parameter representation, the black
disc limit and unitarity}

The elastic amplitude in the impact parameter representation in our
normalization is
\begin{equation}
h(s,b)\ = \ {1\over 2 s} \int^\infty _0 dq\ q J_0(b q)
A(s,-q^2)\ , \quad  \ q=\sqrt{-t}\ .
\end{equation}

The impact parameter representation for linear trajectories
\footnote {Other cases were treated \eg in \cite{VJS}. }
is calculable explicitly for the DP model (13)
\begin{equation}
h(s,b)\ =i\
g_0\ \big[ e^{r_1^2\delta} \ e^{-{b^2 /4 R_1^2}}\ -\ \epsilon\
e^{r_2^2\delta} \ e^{-{b^2 / 4 R_2^2}}\big]\ ,
\end{equation}
where
\begin{equation}
R_i^2= \alpha' r_i^2\  \
(i=1,2) \quad \ ;\ g_0= {a\over 4 b_p\alpha' s_0}\ .
\end{equation}
Asymptotically (\ie when $L\gg b_p$, \ie $\sqrt s \gg 2. $ TeV,
with the parameters of Table 1),
\begin{equation} h(s,b)\underarrow{s\to\infty} i\
g(s)\ (1-\epsilon)\ e^{-{b^2\over 4 R^2}} \ ,
\end{equation}
where
\begin{equation}
R^2=\alpha'L \
\quad \ ;\  g(s)=g_0\left({s\over s_0}\right)^\delta \  .
\end{equation}

To illustrate the $s$-channel unitarity,
we display in Fig.~1 a family of curves showing the imaginary part of
the amplitude in the impact parameter-representation at
various energies; also shown is the calculated (from (1)) inelastic
overlap function.

\medskip
\begin{center}
\includegraphics*[scale=0.9]{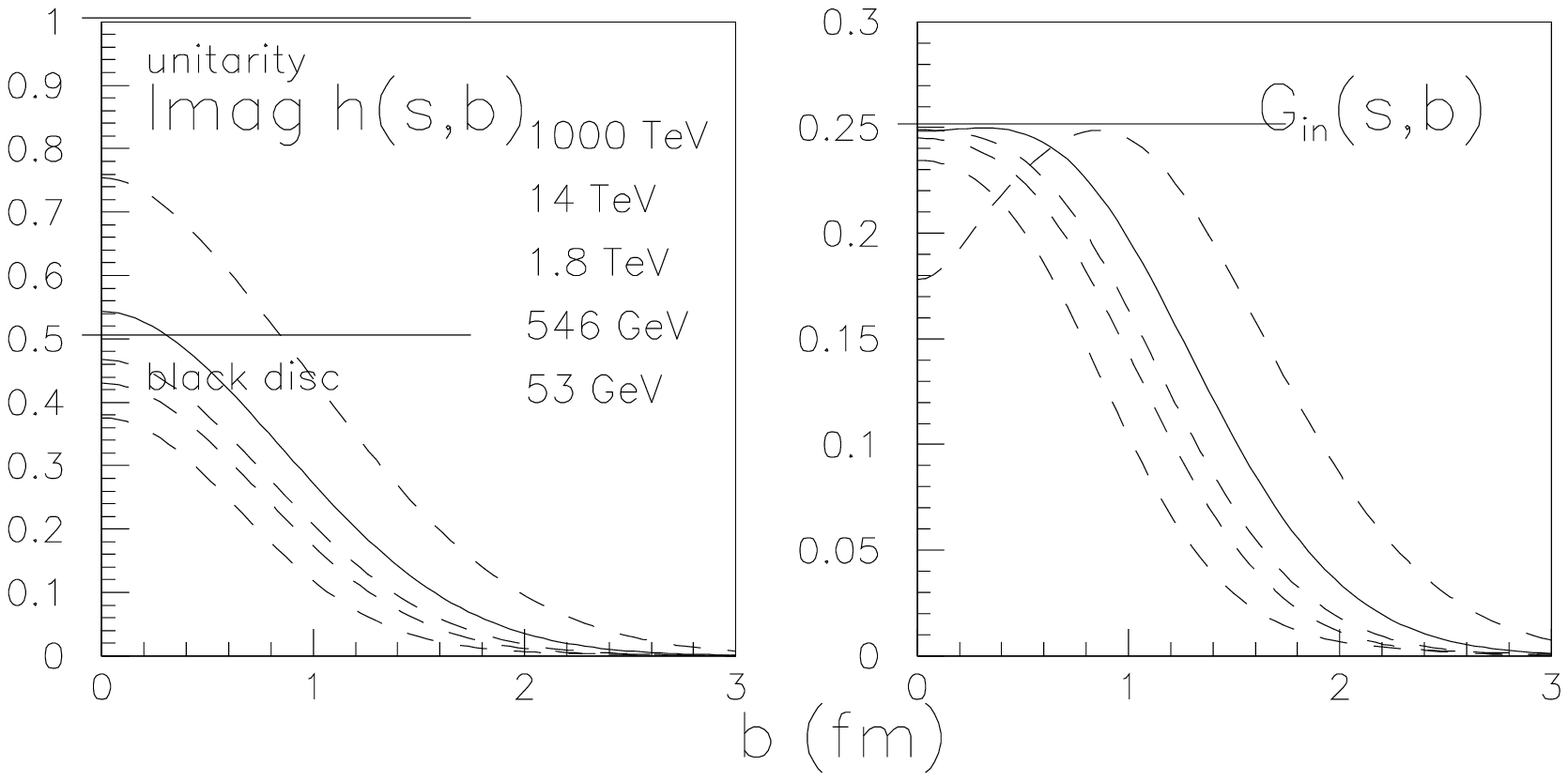}
\end{center}

{\bf Fig.~1.} Calculated "Born level" $\Imag h(s,b) $
and $G_{in}(s,b)$ plotted versus the modulus of the impact parameter
$b$ for some characteristic energies $\sqrt{s}$ as indicated
(solid curve for the
LHC energy). The top of the scale on the left is the unitarity limit
and the value 1/2 corresponds to the black disc limit.  The
calculations are performed for the dipole Pomeron model;
similar results are obtained for the D-L model, see the text.

\medskip

Our confidence in the extrapolation
of $\Imag h(s,b)$ to the highest energies
rests partly on the good agreement of our (non fitted) results with 
the experimental analysis of the central opacity of the nucleon  (see 
Table 2).

\begin{table}
\caption{Central opacity of
the nucleon $\Imag h(s,0)$ calculated at ISR, SPS, Tevatron energies compared
with experiment.}

\medskip
\begin{center}
\begin{tabular}{|c|c|c|c|} \hline
$\sqrt{s}$ & 53 GeV & 546 GeV  & 1800 GeV \\
\hline
exp& 0.36\cite{CS} & $0.420\pm 0.004$ \cite{Ab}& $0.492\pm 0.008$ \cite{Gi} \\
th& 0.36 & 0.424 & 0.461 \\
\hline
\end{tabular}
\end{center}
\end{table}

It is important to note that the unitarity bound 1 for
$\Imag h(s,b)$ will not be reached at the LHC energy,
while the black disc limit 1/2 will be slightly exceeded, the central
opacity of the nucleon being $\Im{\rm m}h(s,0)=0.54$.

The black disc limit is reached at $\sqrt{s}\sim 2 $ TeV, where the overlap
function reaches its maximum ${1\over 4}$. This energy corresponds to
the appearance of the antishadow mode in agreement with the general
considerations in ~\cite{TT}. 
Notice that while $\Imag h(s,b)$ remains central all the way,
$G_{in}(s,b)$ is getting more peripheral as the energy increases starting from 
the Tevatron. For example at $\sqrt{s}=14$ TeV, the central region of 
the antishadowing mode below $b\sim 0.4$ fm is discernible from 
the peripheral region of shadowing scattering beyond $b\sim 0.4$ fm, where 
$G_{in}(s,b)={1\over 4}$.
In terms of \cite{BEL}, the proton will tend to become more
transparent at the center ("gray", in the sense of becoming a gray object 
surrounded by a 
black ring), \ie it is expected to become "GEL" instead of "BEL".

The $s$ channel unitarity limit will not be endangered until
extremely high energies ($10^5$ for the D-L model and $10^6$ GeV
for the DP), safe for any credible experiment. It is interesting to
compare these limit with the limitations imposed by the
Froissart-Martin bound:  actually the Pomeron amplitude saturates the
F-M bound at $10^{27}$ GeV. As expected, the F-M bound is even more
conservative than that following from $s$-channel unitarity.

The D-L and DL models are confronted in the Appendix.

\bigskip

\section{Unitarization}

Now, we consider the unitarized amplitude according to
the "$U$-matrix" prescription \cite{umatrix,VJS}
\begin{equation}
H(s,b)={h(s,b)\over{1-ih(s,b)}},
\end{equation}
with the "Born term" $h(s,b)$ defined in the previous section in (13)-(14).

Fig.~2 shows the behavior of the unitarized impact parameter
amplitude $H(s,b)$ and the corresponding
inelastic overlap function at various energies. By
comparing it with similar curves (Fig.~1) obtained at the "Born level" we
see that unitarization lowers significantly both the elastic and
inelastic impact parameter amplitudes.
\begin{center}
\includegraphics*[scale=0.9]{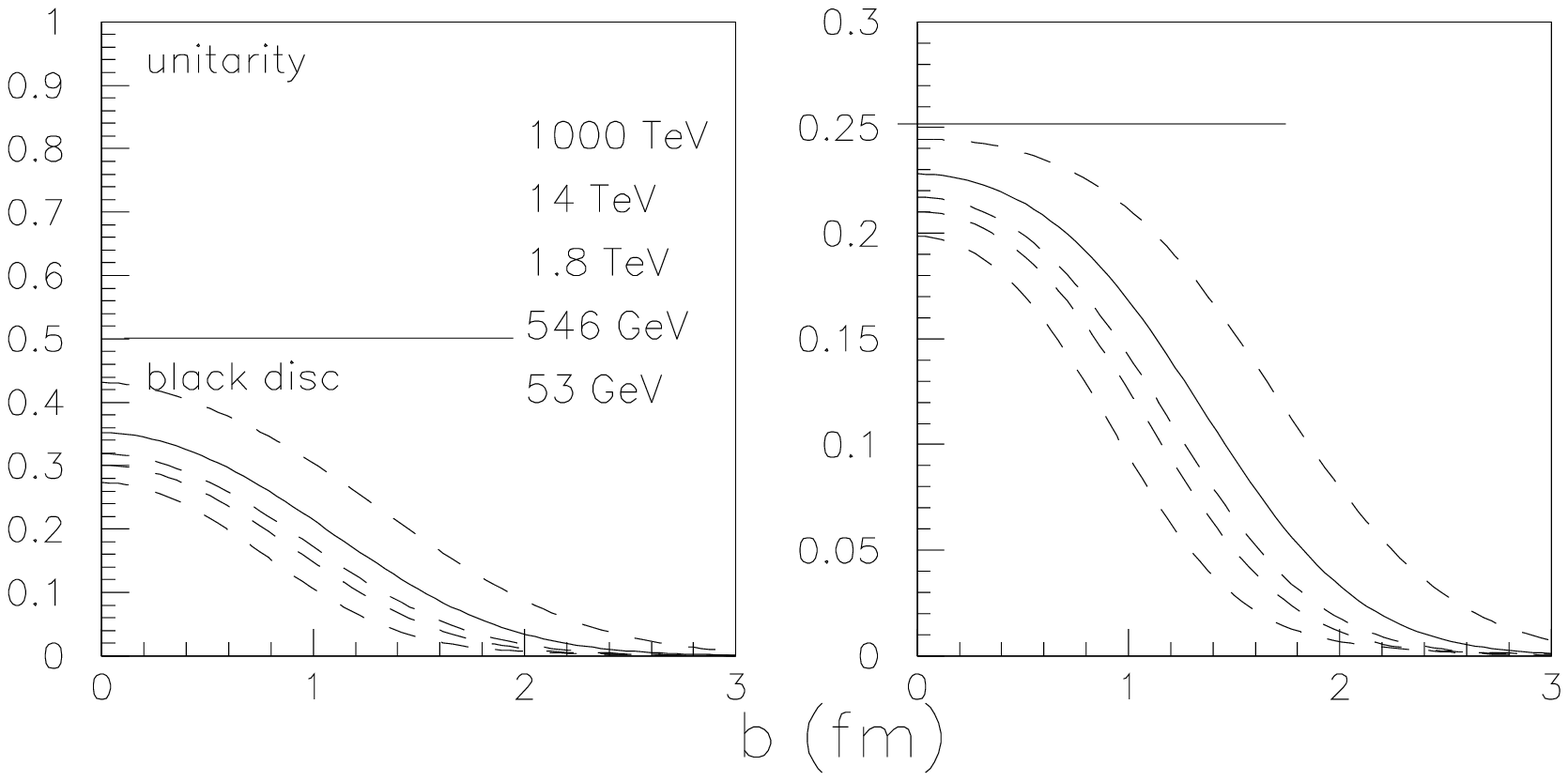}
\end{center}

{\bf Fig.~2.} Same as in Fig.~1, for the unitarized amplitude
$H(s,b)$ and the overlap function, calculated without
refitting the parameters used at the Born level.

An unescapable consequence of the unitarization is that, when calculating the
observables, one should also replace
the Born amplitude $A(s,t)$ by a unitarized amplitude $\widetilde {A}(s,t)$
defined as the inverse Fourier-Bessel transform of $h(s,b)$
\begin{equation}
\widetilde {A}(s,t)\ = \ {2 s} \int^\infty _0 db\ b J_0(b \sqrt{-t})
H(s,b)\ .
\end{equation}
Thus, the above picture may change since the parameters
of the model should in principle be refitted under the unitarization procedure
(this effect of changing the parameters was clearly demonstrated \eg in
\cite{CGM}).

Actually, searching for a new fit of the parameters using an
unitarization procedure is  time consuming and unnecessary for the
present discussion because
the behavior of the amplitude and of the overlap function in the impact
parameter representation obtained at the Born level will be almost restored 
after unitarization.  We checked that the parameters of the complete 
model (with the secondary Reggeons and Odderon added in the fit) after 
unitarization may be rearranged so as to  
reproduce well the data and give roughly the same extrapolated properties 
as at the Born level.

While the unitarity limit now
is secured automatically (remind that $\Imag h(s,0)$ is well below 
that limit even at the "Born level" in the TeV region) 
the behaviour of the elastic impact parameter 
amplitude after it has reached the black disc limit corresponds (see 
\cite{TT}) to the transition from shadowing to antishadowing. In 
other words, the proton (antiproton) after having reached its maximal 
blackness around $2$ TeV, will become gradually more transparent 
with increasing energies at its center.

\medskip

\bigskip

\section{Conclusions}

While the results of our analyzis in the impact parameter representation are
in agreement with the earlier observations that
$\Imag{h(s,b)}$ remains central and $G_{in}(s,b)$ 
becomes peripheral as the energy exceeds 2 TeV 
(see Fig.~1). There is a substantial difference with the known
"BEL-picture" \cite{BEL}, according to which with increasing energy
the proton becomes Blacker, Edgier and Larger.

We confirm that getting edgier and larger, the proton, after reaching 
its maximal blackness, will tend to be more transparent or ("GEL",
\ie a gray disc surrounded by a black ring) 
when the energy exceeds that of the Tevatron. This transition
from shadowing to new antishadowing scattering mode is expected to  
occur at the LHC. 

To conclude, we stress once more that both the data and relevant
models at present energies are well below the $s$-channel unitarity
limit. In our opinion, deviations due to the diversity of
realistic models may result in discrepancies concerning $\Imag{h(s,0)}$
of at most $10\%$, while its value at 2 TeV is still half that of the
unitarity limit, so there is no reason to worry about it!  Opposite
statements may result from confusion with normalization. Therefore,
model amplitudes at the "Born level" may still be quite interesting
and efficient in analyzing the data at present accelerator energies
and giving some predictions beyond. The question, which model is
closer to reality and meets better the requirements of the
"fundamental theory" remains of course topical.

Extrapolations and predictions  to the
energies of the future accelerators (see e.g. \cite{Chatr}) are both
useful and exciting since they will be checked in the not-so-far
future at LHC and other machines. The fate of the "black disc limit"
is one among these.

\bigskip

\noindent
\large{{\bf Acknowledgements}} We thank S.M. Troshin for a useful 
correspondance.

\bigskip

\begin{center}
\large {APPENDIX}

{\bf Comparaison between the DP and D-L models}
\end{center}

The D-L amplitude in the impact parameter representation at the
Born level, calculated from (9) and (17) is
\begin{equation}
h(s,b)=-{N\over 2s}\left(-i{s\over s_{dl}}\right)^{\alpha(0)}
{e^{-b^2\over 4B'(s)} \over 2B'(s)}\ ,
\quad B'(s)= B+\alpha'(\ell n{s\over s_{dl}}-i{\pi\over 2})\ .
\end{equation}
As already noted, the $s$ channel unitarity limit both for the DP and 
the D-L model will not be
endangered until extremely high energies ($10^5$ GeV for the DP and
$10^6$ GeV for the DP model, the order-of-magnitude differences coming
from the smaller intercept in the DP model), while the F-M bound is 
saturated at $10^{27}$ GeV (for more details see \cite{BDI}).

Table 3 presents a selection of results concerning the DP and D-L
models for the Pomeron in the impact parameter
representation of the elastic amplitude and inelastic overlap
function, calculated at $b=0$ at the LHC energy.

We conclude that the two models give similar results; all conclusions
on unitarity and black disc limits for DP model hold for
D-L model as well (the curves in Figs.~1,2 would be
indistinguishible by eye).

Note that both  models are supercritical,
with asymptotic $s^\delta$ type behavior of the total cross sections.
They are known to give fits
which cannot be discriminated by present data
from an asymptotic $\ell n^2s$ type behaviour.
This is another argument to neglect unitarization effects.

\begin{table}
\caption{Maximum values of the amplitude and overlap function at the
Born level and after $U$-matrix unitarization calculated at 14 TeV
for the DP and D-L models without refitting the parameters}
\medskip
\begin{center}
\begin{tabular}{|c|c|c|c|c|} \hline & $\Imag h(s,0)$ &
$G_{in}(s,0)$ &   $ \Imag H(s,0)$ & $\widetilde{G_{in}}(s,0)$\\
\hline
DP& 0.535  & 0.247 & 0.349 & 0.227\\
D-L& 0.539 & 0.246 & 0.351 & 0.227\\
\hline
\end{tabular}
\end{center}
\end{table}

\end{document}